# Demodulation of Spatial Carrier Images: Performance Analysis of Several Algorithms Using a Single Image

C. Badulescu, M. Bornert, J.-C. Dupré, S. Equis, M. Grédiac, J. Molimard, P. Picart, R. Rotinat, V. Valle

C. Badulescu

LBMS, EA 4325, ENIB-ENSTA-UBO,

Brest, France

M. Bornert

Laboratoire Navier, UMR 8205, CNRS-École des Ponts ParisTech,

Champs-sur-Marne, France

J.-C. Dupré • V. Valle

Institut P', UPR 3346, CNRS-Université de Poitiers-ENSMA,

Futuroscope Chasseneuil, France

S. Equis

EPFL, Nanophotonics and Metrology Laboratory,

Lausanne, Switzerland

M. Grédiac (SEM Member)

Institut Pascal, UMR 6602, Université Blaise Pascal-CNRS,

Clermont-Ferrand, France

J. Molimard

LGF, UMR 5307, Ecole Nationale Supérieure des Mines, CIS-EMSE, CNRS,

Saint-Etienne, France

P. Picart

LAUM, UMR 6613, CNRS-Université du Maine,

Le Mans, France

R. Rotinat (SEM Member)

LMPF, EA 4106, Arts et Métiers ParisTech

Châlons-en-Champagne, France

**Abstract.** Optical full-field techniques have a great importance in modern experimental mechanics. Even if they are reasonably spread among the university laboratories, their diffusion in industrial companies remains very narrow for several reasons, especially a lack of metrological performance assessment. A full-field measurement can be characterized by its resolution, bias, measuring range, and by a specific quantity, the spatial resolution. The present paper proposes an original procedure to estimate in one single step the resolution, bias and spatial resolution for a given operator (decoding algorithms such as image correlation, low-pass filters, derivation tools ...). This procedure is based on the construction of a particular multi-frequential field, and a Bode diagram representation of the results. This analysis is applied to various phase demodulating algorithms suited to estimate in-plane displacements.

**Keywords** Fringe • Error assessment • Spatial resolution • Displacement resolution • Benchmark

# 1 Introduction

Full-field optical methods are commonly used in experimental mechanics to obtain displacement and strain fields. Most of these methods are powerful tools which are now easy to use, but their metrological performances are neither clearly defined nor completely characterized. In particular, there is no standard available to calibrate these measurement systems. This lack of information undoubtedly leads to limit their use in industry, especially when quantitative and reliable information is needed to characterize the mechanical response of materials or structures.

A recent EC-funded project named SPOTS [1] proposed a procedure suitable to calibrate various full-field measurement systems [2, 3]. This procedure is based on the use of a testing device and a specific specimen for which the displacement/strain/stress distributions are assumed to be reliably known, thus leading to a reference for the systems to be calibrated. This project has not (yet) turned into a standard.

Another approach proposed by the French network of research groups GDR CNRS 2519 devoted to full-field measurement techniques and named "Full-field measurement methods and identification in solid mechanics" [4], consists in generating synthetic images containing a reference displacement and in extracting this displacement with the software to be tested. Comparing this synthetic "measured" displacement field and the reference one enables one to assess the performance of the algorithm. This type of approach has been recently proposed to assess the metrological performance of various Digital Image Correlation (DIC) packages [5]. Several series of pairs of synthetic images of random patterns (reference and deformed ones) were generated. The displacement used to construct the deformed images was assumed to be sinusoidal. These images were processed by various DIC programs. This led to observe some interesting trends in the obtained results, which do not directly depend on the programs themselves, but on the underlying algorithms.

DIC is a technique which has recently widely spread in the experimental solid mechanics community, but there are also some very popular techniques in which the information is encoded in the phase modulation of a temporal or spatial carrier and not in a random pattern. The most popular techniques based on carrier modulation encoding are moiré, moiré interferometry, speckle interferometry and the grid method [6-7]. In these cases, fringes are processed and the sought displacement is contained in the phase of the signal. Many phase-extraction algorithms have been proposed in the literature [8-9] but there is no general study aimed at assessing the performance of these algorithms, which could be considered as equivalent to that proposed for DIC in [5]. A thorough investigation of various procedures used to extract phases from fringe patterns is proposed in [10], but no simulated images are processed in this study.

In this context, the aim of this paper is to examine the metrological performances of various algorithms developed by some of GDR CNRS 2519 partners. These programs are suitable for phase extraction from a single fringe patterns. Consequently, they should be used especially with the grid method, but also to many high speed implementations (for example: fringe projection, deflectometry or speckle or moiré interferometry).

The proposed methodology is based on a generic synthetic image undergoing sinusoidal displacements. The procedure used to generate original images containing very rich information in terms of frequency and phase distributions is presented and justified in the first section of this paper. Displacements are computed by seven fringe analysis codes for various parameters. The measured displacements are compared to the prescribed ones which serve as a reference through a specific method which is described. The two main interests are *i-* to rely on a reference phase distribution which is perfectly known *a priori* and *ii-* to control any additive noise that can potentially be added to this reference, thus leading to a precise assessment of its influence on the measurements provided by various algorithms. Obtained results are presented, analyzed and discussed.

## 2 Principle of error assessment

*Introduction*

The performance of a given algorithm depends on various properties, among which its ability to detect localized phase variations. This type of information is somewhat difficult to be taken into account in a systematic way in a synthetic image. However, it is well-known that, when using a spatial Fourier transform, any arbitrary displacement can be expressed as a sum of single spatial frequency functions, each of them featuring a given amplitude, frequency, phase and direction. Consequently, the objective here is to examine how such functions are identified by the algorithms under study, as already proposed for DIC [5]. Rather than examining separately various reference functions, each of them exhibiting given frequencies and phases, this approach proposed to construct an image in which these quantities could spatially change. The main advantage is that processing only one image provides very rich information concerning the influence of these parameters

(frequency, phase) on the obtained result. The construction of this image is detailed in the following section.

*From phase to displacements*

The results of phase extraction algorithms are either a phase or a phase variation. In common cases, this quantity can be related to mechanical information, depending on the nature of the set-up and of the fringe generation. If considering only the geometrical techniques (Fig. 1), the fringes may be projected onto the surface; the diffused light image, and consequently its phase, contains the shape of the specimen (fringe projection technique [11, 12]); the reflected light contains information on the slope of the surface (deflectometry [13]). For the grid method [14], the "fringes" are directly represented by the lines of a grid transferred or engraved on the surface of the specimen under study; the information encoded in the phase map is the in-plane displacement. Last, some interferometry techniques also lead to fringe patterns with in-plane or out-of-plane sensitivity, such as speckle [15~17] or moiré interferometry [18], digital holographic interferometry [19, 20] (displacement), shearography [21] (displacement gradients).

{Figure 1 might be around here}

A general model for fringe images encoded with a spatial-carrier frequency has been described for example in [9] or [22]. The measurand (i.e. the displacement in the case of grid technique) is linked (linearly or not) to the phase of the signal. Let us consider for example a unidirectional spatial carrier; $y$ is one of the directions of the reference set of coordinates attached to the surface of the specimen under study. Any point $C$ of the sensor of the camera corresponds to a physical point $M$ of the object, denoted $M_0$ in the reference configuration and $M_1$ in the current configuration. The light distribution modeling the fringes can be modeled as follows:

$$I(C) = I(M_0) = I_0 \left[1 + \gamma frng\left(2\pi \vec{f}_y \cdot \vec{r}_0\right)\right] \tag{1}$$

where $I_0$ is the average intensity, $\gamma$ ($0<\gamma<\square$) is the contrast or visibility of the signal, *frng* is a $2\pi$ periodic continuous function (i.e. cosine function in case of harmonic pattern), $\vec{f}_y$ is the spatial-carrier frequency vector associated to the fringes. Its components in the $(x,y)$ basis are $(0, f_0)$, with $f_0=1/P_{car}$ ($P_{car}$: carrier period).

When the object is deformed, in the current configuration, $M_0$ becomes $M_1$. $M_1$ is characterized, in the reference configuration, by a position vector $\vec{r}_1(x, y)$ in the Cartesian reference basis. The intensity at point $M_1$ can be written as follows:

$$I(M) = I(M_1) = I_0 \left[1 + \gamma frng\left(2\pi \vec{f}_y \cdot \vec{r}_1\right)\right] \tag{2}$$

However, the deformation from the reference state to the current state is mathematically described by the displacement field $\vec{U}(M_1) = \vec{r}_0 - \vec{r}_1$. Hence Eq. 2 becomes:

$$I(M_1) = I_0 \left[1 + \gamma frng\left(2\pi \vec{f}_y \cdot \left(\vec{r}_0 - \vec{U}(M_1)\right)\right)\right] \tag{3}$$

If the deformation is small between two images, displacements $\vec{U}(M_1)$ and $\vec{U}(M_0)$ are almost equal. From the reference state to the current state, the phase $\varphi_y = 2\pi \vec{f}_y \left[\vec{r}_0 - \vec{U}(M_0)\right]$ of the *frng* function evolves in such a way that $\Delta\varphi_y = -2\pi \vec{f}_y \cdot \vec{U}(M_0)$.

*Modeling the displacement field*

The current work is based on the analysis of synthetic fringes which were created from a known displacement field $U_y$. Analyzing a periodical signal $U_y$ is a common way to characterize a signal processing algorithm, as already used in [5] for instance. The main problem in this previous work was the great number of images which had to be generated and processed. In order to solve this problem, multiple frequency responses are analyzed with only two images: a reference one and a deformed one. Of course, frequencies must be clearly identified in the displacement field. For the deformed image, the use of a periodic displacement featuring a variable period along the *y*-axis is proposed. This displacement can be written as follows:

$$U_y = A_U \cos(F) \quad \text{with} \quad F = \frac{\pi}{P} \frac{y^2}{y_{max}} \quad (4)$$

where $A_U$ is a constant amplitude, $y_{max}$ is the image height and $P$ is the minimal wavelength of the periodic displacement. This choice is justified by the fact that it leads to a linear evolution of the local frequency according to *y*:

$$f_y = \frac{1}{2\pi} \frac{\partial F}{\partial y} = \frac{1}{P} \frac{y}{y_{max}} \quad (5)$$

This leads to a wide frequency range adjusted by the parameters $y_{max}$ and $P$. It can be noted that the minimal wave length of the created periodic displacement is reached at the bottom of the image.

As mentioned above, the idea is to investigate with a unique image all the possible frequencies. The local optical phase should add some specific errors, such as, for example the presence of harmonics [23]. So as to separate the effect of the phase from that of the frequency, a term corresponding to a linear evolution of the displacement phase along the *x*-axis was added. The displacement can eventually be written as follows:

$$U_y = A_U \cos\left(\frac{\pi}{P}\frac{y^2}{y_{max}} + 2\pi\frac{x}{x_{max}}\right) \qquad (6)$$

Fig. 2 illustrates the displacement field obtained with $A_U$=0.3, $P$=5, $y_{max}$=1024 and $x_{max}$=512 (dimensions are in pixels).

{Figure 2 might be around here}

*Generation of synthetic images*

As the goal is to obtain spatially-modulated image fringes, image encoding can be idealized as a periodic pattern. In this paper, a perfect sinusoidal modulation is used in the first round robin test between various algorithms. Eq. 3 leads to:

$$I = I_0\left[1 + \gamma\cos\left(2\pi\frac{y}{P_{car}} + 2\pi\frac{U_y}{P_{car}}\right)\right] \qquad (7)$$

More complex patterns can also potentially be generated, such as triangular or rectangular ones, only by adding harmonic frequencies. For instance, a rectangular pattern will be studied below by using 60 terms. It is worth noting that the main advantage of this analytical formulation is to introduce no bias. Indeed, any approximated solution would have relied on interpolation procedures, and this would have led to some bias.

Image acquisition inevitably induces some differences between ideal and actual patterns. One of these differences is due to encoding depth. This phenomenon is simulated here by using 8-bits encoding. In the same way, random Gaussian noise can also be added to the images.

The actual images are also digitized as is done with the CCD sensor of a camera; practically, each sensor element acts as a spatial integrator of the image intensity. Hence, perfect sine-wave and perfect square signals are deformed and this effect has also been taken into account. In order to achieve an $x_{max} \times y_{max}$ final image, an intermediate image which is $k$ times over-sampled along each direction is generated. Then, a $k \times k$ moving-average filter is applied to the intermediate image. The last step is to under-sample this intermediate image to get the final $x_{max} \times y_{max}$ image, which is considered as input data for the algorithms tested in this study.

Some other effects such as lighting variations or diffraction spots might be analyzed in the same way but these issues are not addressed in this study for the sake of simplicity.

The parameters used to generate the images are summarized in Table 1. The reference image is built with $U_y=0$ and the deformed one with the expression given by Equation (3). The following parameters are fixed:

- the image size ($x_{max}$, $y_{max}$),
- the minimum displacement wavelength ($P$),
- the displacement amplitude ($A_U$),
- the mean intensity and modulation ($I_0$, $\gamma$),
- the spatial integration of pixels (over-sampling),
- the encoding depth.

In Table 1 and Table 2, Case 1 represents a reference case, with a perfect sine wave, no noise, and parameters which are equal to the typical values characterizing actual sensors. The effects of both the noise and the shape of the *frng* function are considered separately in Cases 2 and 3. Case 4 is equivalent to Case 3 but with an additional noise. Last, check tests were carried out to observe and quantify the influence of the displacement amplitude (Cases 5 and 6) and the over-sampling on the identified displacement (Case 7).

{Table 1 and 2 might be around here}

# 3 Presentation of the different fringe analysis methods

The information characterizing the simulated displacement fields is encoded in the phase modulation of a spatial carrier. Among the various approaches available in the literature to demodulate this signal, one can distinguish four different types which are used in practice by the GDR2519 partners: by using a defined transformation, by phase-shifting, by morphological analysis or with the DIC technique.

*Fourier-transform techniques*

*Spatial Fourier-Transform (SFT-LAUM)*

The spatial Fourier-transform was proposed first by Takeda [24] and became a very popular method for fringe demodulation [25], in particular for surface profilometry [26]. In the Fourier spectrum, the spatially modulated fringe image is represented with three orders: the first one is localized at the center of the spatial spectrum, the second is localized at spatial frequency $+1/P_{car}$, and the last one is localized at spatial frequency $1/P_{car}$, $P_{car}$ being the pitch of the spatial carrier.

The fringe image recorded by the pixel matrix of the sensor can be simply written as:

$$I(x,y) = I_0(x,y)\left[1 + \gamma\cos\left(\varphi_d(x,y) + 2\pi\frac{y}{P_{car}}\right)\right] \quad (8)$$

Developing Eq. 8 leads to:

$$I(x, y) = \frac{1}{2} \gamma I_0(x, y) \exp(i\varphi_d(x, y)) \exp\left(2i\pi \frac{y}{P_{car}}\right)$$
$$+ \frac{1}{2} \gamma I_0(x, y) \exp(-i\varphi_d(x, y)) \exp\left(-2i\pi \frac{y}{P_{car}}\right) + I_0(x, y) \quad (9)$$

Denoting:

$$c(x, y) = \frac{\gamma}{2} I_0(x, y) e^{i\varphi_d(x,y)} \quad (10)$$

and applying the spatial Fourier transform to Eq. 8 leads to:

$$\tilde{I}(u, v) = A(u, v) + C\left(u, v - \frac{1}{P_{car}}\right) + C^*\left(-u, -v - \frac{1}{P_{car}}\right) \quad (11)$$

with

$$A(u, v) = FT[I_0(x, y)](u, v) \quad (12)$$

and

$$C(u, v) = FT[c(x, y)](u, v) \quad (13)$$

where FT means Fourier Transform.

If the spatial frequency spectrum of the three orders is not expanded with "too many" terms, the three orders are well separated, so they can be extracted using a filtering mask. Denoting $f[...]$ the filtering in the Fourier spectrum, the inverse Fourier transform applied to the filtered spectrum can be written as follows:

$$\hat{c}(x, y) = \frac{\gamma}{2} I_0(x, y) \exp[i\varphi_d(x, y)] \exp\left(2i\pi \frac{y}{P_{car}}\right)$$
$$= FT^{-1}\{f[FT[I(x, y)]]\} \quad (14)$$

For example, the filtering $f[…]$ can be binary rectangular centered at spatial frequency $(0; 1/P_{car})$ with width $\Delta u \times \Delta v$ pixels$^{-2}$.

The phase can be extracted using an arc tangent formula:

$$\varphi_1(x, y) = \arctan_{2\pi}\left(\frac{\Im_e(\hat{c}(x, y))}{\Re_e(\hat{c}(x, y))}\right) \qquad (15)$$

$$= \varphi_d(x, y) + 2\pi \frac{y}{P_{car}}$$

Where $\arctan_{2\pi}$ corresponds to the four-quadrant inverse tangent defined in the interval $[-\pi,\pi]$.

In the reference state, the demodulated phase $\varphi_d$ represents a constant phase difference between the fringe carrier and the real experimental state, summing up experimental set-up errors (material alignment, grid defects ...) or phase extraction assumption (pitch error due to discretization). All these errors are supposed to be constant during the experiment. So far, in the initial state,

$$\varphi_d(x, y) = \varepsilon(x, y) \qquad (16)$$

When the object is deformed, the extracted phase becomes:

$$\varphi_2(x, y) = \varepsilon(x, y) + 2\pi \frac{y}{P_{car}} + \Delta\varphi(x, y) \qquad (17)$$

Then, the phase change due to the simulated specimen deformation is calculated with $\Delta\varphi = \varphi_2 - \varphi_1$. The spectral bandwidth ranges from a cut-off frequency of $f_{min}=0.9/P_{car}$ up to the highest frequency involved in the discretized signal ($f_{max}=1/2$).

*Spectral Analysis (SA-Institut P')*

This spectral analysis method is based on the calculation of a local spatial Fourier transform (Eq. 11) of a sinusoidal signal (Eq. 8) on subsets with a size of $M \times N$ pixels (M lines and N columns). As in the SFT technique, the spectrum can be decomposed in three peaks: assuming that the signal frequency is constant over the subset, the second peak in the frequency spectrum is a function of the characteristics of the sinusoidal signal. Its location is given by the number of fringes along the horizontal and vertical directions. The process of calculation is the same for the two directions. For the vertical direction (i.e. position in y direction) the number of fringes is given by: $K=M/P_{car}$. As a discrete Fourier transform is used, $K$ is an integer value, so an interpolation process is used to obtain a fractional part $\beta$ [26] in order to obtain an accurate value of the pitch:

$$P_{car}=M/(K+\beta) \text{ with } -0.5<\beta<0.5 \tag{18}$$

$\beta$ is defined by [27][28]:

$$\beta = \Re_e\left(\frac{C_{K-1} - C_{K+1}}{2C_K - C_{K+1} - C_{K-1}}\right) \tag{19}$$

with $C_K$ the complex value of the Fourier-transform of the signal for the peak of coordinate $K$ and $C_{K-1}$, $C_{K+1}$ the adjoining complex values of the peak $K$. The same process is applied in the x direction in order to obtain the pitch and orientation of the grating [28]. Then this procedure can be applied for the study of various materials with natural grating like textile [28], or with artificial grating like paper [28, 29] or single lap joint [30].

The complex amplitude and the reference phase are given by:

$$A = \frac{j2\pi\beta(\beta^2 - 1)}{1 - e^{j2\pi\beta}}(2C_K - C_{K-1} - C_{K+1}) \tag{20}$$

$$\varphi_1 = \arctan_{2\pi} \frac{\Im_e(A)}{\Re_e(A)} \tag{21}$$

The reference phase on a subset may be obtained by this last expression. The whole phase fields in the reference state and in the current state are obtained by shifting the subset. Eq. 17 is used to obtain the phase change.

The spectral bandwidth limits, in this case, range from $K$-1 to $K$+1, i.e. $f_{max}$-$f_{min}$=2/$M$.

For this study, the sizes of the computation subset are respectively 32x32 and 16x16 pixels, commonly used, and 32x8 pixels, an optimized subset taking into account the geometry of the studied gratings.

*Spatial Phase-Shifting method (SPS-IP, LGF, LMPF)*

Eq. 1 is the general form of a measured intensity field. Without any assumption on the *frng* function, there are an infinity of unknown parameters: among them, of course, the phase, which will have to be identified. Thus, the phase cannot be deduced by recording the intensity field only, since there is one equation for at least four unknowns: the mean intensity $I_0$, the contrast $\gamma$, the fringe pitch $P_{car}$ and the phase $\varphi$ (if function *frng* is harmonic). This is the reason why supplementary equations are required to solve the system. These additional equations can be obtained by two different methods of phase shifting, which are:

– spatial phase-shifting: the information is searched at the vicinity of the considered pixel. This approach is used especially when there is a regular dense carrier fringe, as in the grid technique.

– temporal phase shifting: the information is searched at the considered pixel from a set of images obtained with a known phase shift. This method is used with interferometry or fringe projection set-ups, leading to very high spatial resolution.

The first technique is chosen here because the idea is to extract displacement (i.e. phase variation) from only one fringe image. The technique used here is based on a *N*-bucket algorithm [14]. The number of pixels sampling one period of the fringe signal is an integer denoted *N*. The principle is to get *M* intensity samples noted $I_k$, with $k=0,1,\ldots,M$-1, separated by a constant phase shift $\delta=2\pi/N$:

$$I_k = I(\varphi + k \cdot \delta) \tag{22}$$

These sampling points enable one to determine the best fitting harmonic function. With this approach, the general form of the phase detection algorithm is given by the following equation:

$$\varphi = \arctan_{2\pi}\left[\frac{\sum_{k=0}^{M-1} b_k I_k}{\sum_{k=0}^{M-1} a_k I_k}\right] \quad (23)$$

The quality of the phase strongly depends on the way the $a_k$ and $b_k$ coefficients are chosen. By choosing a windowed discrete Fourier transform (W-DFT) algorithm with a triangular windowing, harmonics are eliminated up to the $(N-2)^{th}$ order and the influence of the uncertainty on the calibration (number of pixels per period i.e., $N$) is reduced [14].

The phase extraction is performed by seeking the particular term argument of the local Fourier-transform of the intensity map. It is necessary to define an arbitrary pitch to demodulate the signal of the phase. This arbitrary pitch is obviously different from the effective grid pitch, because it is difficult to determine it precisely in a real experiment, and because the grid pitch is effectively different in the reference and the current states. In this case, for a given set of $2N-1$ pixels, the phase is defined by:

$$\varphi = -\arctan_{2\pi} \frac{\sum_{k=1}^{N-1} k(I_{k-1} - I_{2N-k-1})\sin(2k\pi/N)}{NI_{N-1} + \sum_{k=1}^{N-1} k(I_{k-1} - I_{2N-k-1})\cos(2k\pi/N)} \quad (24)$$

where $k$ is the coordinate of the current point and $I_k$ is the local intensity of the image. This formulation is in fact a mere convolution between a Fourier-transform of the signal and a triangular window. The implementation used in laboratories [31~33] supposes an adaptation of the coding pitch to each situation (5-6-7 or 8 pixels per pitch). The width of the bi-triangular window used here is $2N$-1 pixels, i.e.: 9, 11, 13 or 15 pixels respectively.

Last, even if the principle of the technique is the same among the three groups using this technique, results might be different for the same technique because of the implementation. Some valuable examples using the SPS should be found for concrete [34], steel [35] or composite materials [36].

*Correlation techniques*

*Modulated Phase Correlation (MPC-Institut P')*

The principle of the Modulated Phase Correlation algorithm (MPC) lies in the adaptation of the correlation technique. In a zone of interest, it estimates the degree of similarity between a real fringe pattern and a mathematical model [37-38].

From Eq. 1, a mathematical representation of a two-dimensional fringe pattern can be globally expressed by:

$$I_v(x,y) = I_0(x,y)(1 + \gamma(x,y)\cos[\varphi(x,y)]) \qquad (25)$$

where $I_0(x,y)$ is the amplitude modulation, $\gamma(x,y)$ the fringe contrast, and $\varphi(x,y)$ the phase value. In this global representation $I_0$, $\gamma$ and $\varphi$ are complicated functions that depend on coordinates $x$ and $y$. If a smaller zone of interest centered at coordinates $(x,y)$ is considered, the global expression can be reduced to a simpler one in which $\varphi$ depends on local coordinates $(\xi,\gamma)$, and where $x$ and $y$ are fixed. The current mathematical model representing fringes with constant pitch and orientation is described by:

$$I_0(1 + \gamma\cos[\varphi(\theta, p_{car}, \varphi_d, \xi, \gamma)]) \qquad (26)$$

with:

$$\varphi(\theta, p_{car}, \varphi_y, \xi, \gamma) = \frac{2\pi}{p_{car}}\cos(\theta)(x+\xi) + \frac{2\pi}{p_{car}}\sin(\theta)(y+\gamma) + \varphi_y \qquad (27)$$

Variables $P_{car}$ and $\theta$ are respectively the pitch and orientation of the fringe pattern and $\varphi_d$ is the term of the demodulated phase. Indeed, $I_0$ and $\gamma$ parameters are considered as constant in this zone of interest. Eq. 26 is valid if the zone of interest is limited to a few pixels.

For the calculation of the degree of similarity between the mathematical expression denoted $I$ and the real fringe pattern denoted $I_r$, the following formulation is used at coordinates $(x, y)$:

$$\psi(I_0, \gamma, \theta, p_{car}, \varphi_d) = \iint_D (I_r(x,y) - I(x,y))^2 ds \qquad (28)$$

A gradient method under constraint is employed here to minimize the $\psi$ function because periodic functions that depend on $\alpha$ and $\varphi_d$ are used. The minimum of $\psi$ directly gives the five sought parameters:

$$\psi(I_0^i, \gamma^i, \theta^i, p_{car}^i, \varphi_d^i) = \underset{I_0, \gamma, \theta, p_{car}, \varphi_d}{Min} [\psi(I_0, \gamma, \theta, p_{car}, \varphi_d)] \qquad (29)$$

A discrete formulation of $\psi$ is used in practice. This technique has been employed for the study of material behaviors and for example to determine J-Integrals in fracture mechanics [39]. For the present study, computation is done with zones of interest with dimensions equal to $(8\times 2)$, $(1\times 24)$, $(P_{car}\times 16)$ and $(P_{car}\times 1)$ pixels.

*Digital Image Correlation technique (DIC-Laboratoire Navier)*

Digital Image Correlation (DIC) techniques aim at identifying an approximation of a mechanical transformation linking a reference image to a deformed image, assuming grey level convection by the transformation. Standard implementations split this problem into several independent local ones, defined on small windows usually regularly spread over the whole region of interest. The local mechanical transformation on these windows is based on a local expansion of the displacement field, to some chosen low orders. The purpose of the DIC algorithm is to determine the best parameters of this expansion. They maximize the correlation coefficient measuring the similarity of the grey level distribution in the considered window in the reference image, with that in the deformed image back-convected according to the assumed local transformation. Sub-pixel accuracy can be achieved by means of some interpolations of the grey levels in between the pixels of the deformed image. In practice, one often retains the sole value of the optimal transformation at the center of the windows as an estimate of the displacement at this pixel position of the reference image. This operation can be repeated over all pixels in the reference image as long as the associated windows belong to the image, so as to generate a dense estimate of the displacement field.

Such a procedure has been applied to the present sets of pairs of images, by means of the in-house CMV code developed at Navier, with the following particular choice of parameters. Windows are one pixel wide and 5, 9 and 17 pixels high; the local transformation over these windows is prescribed to be a uniform translation along the vertical direction, characterized by a single value which is the intensity of this translation. The images were interpolated by means of a biquintic interpolation function, using the 6×6 neighboring pixels for a given position. The optimization of the translation parameter was performed with a first-gradient minimizing procedure, which is stopped when the last step is smaller than $0.4 \times 2^{-15} = 10^{-5}$ pixels, which ensures that the accuracy is not limited by the numerical optimization algorithm. The correlation coefficient is a centered normalized cross-correlation.

DIC algorithms are usually applied on images with random patterns, either artificial, like speckle painting, or natural. With a periodic pattern such as those used in so-called grid-methods, there are several optima of the correlation coefficient for the periodically-distributed values of the transformation parameters. In order to deal with such a situation, the DIC implementation has to be slightly adapted to avoid the selection of a wrong local minimum. This can for instance be performed by seeking the local displacement field with a propagation procedure. The detection of the translation components at a new position -- close to one or several already known positions -- is restricted to a domain where there is only one local optimum, the right one. The research domain is defined from the knowledge of the displacement at the neighboring positions -- assuming there is a local continuity of the transformation -- and limited by the periodicity of the pattern. Such a procedure has been successfully applied for years to quantify the deformation of materials at micro-scale by means of micro-grids and *in situ* testing in a scanning electron microscope [29, 37]. It was applied without modification in the present study. It should be noted that with this approach no *a priori* knowledge of the particular shape of the marking is used, and that the intermediate stage of phase retrieval is not required.

In order to fit the conventions used within standard grid processing to define the displacement field – which are adopted in the present study -- the above presented DIC algorithm was applied by using the deformed images as reference images and vice-versa, and retaining the opposite of the DIC displacement field as the quantity to be compared to the prescribed displacements. With standard DIC procedures, the evaluated displacement is $\vec{U}(\vec{r}_0) = \vec{r}_1 - \vec{r}_0$, whereas $\vec{U}(\vec{r}_1) = \vec{r}_1 - \vec{r}_0 = -(\vec{r}_0 - \vec{r}_1)$ is considered in the present study.

## 4. Procedure to analyze the results

Pairs of images were generated using prescribed procedure. Indeed, resulting images are synthetic images, this approach being believed to give a more general point of view on the metrological qualities of image processing algorithms. Synthetic images were analyzed by the seven partners with their own processing tools as described previously: Spectral Analysis (SA), Spatial Fourier Transform (SFT), Spatial Phase-Shifting (SPS), Modulated Phase Correlation (MPC) and Digital

Image Correlation (DIC). Finally results for each processing tool are displacements or phase maps converted into displacements according to the equation:

$$U_y = \frac{\varphi_d}{2\pi} P_{car} \quad (30)$$

For each technique, the results are classified by their method's name (SA, SFT…) following by the sizes of the used subset (vertical and horizontal) and if necessary the grid step.

Displacement fields were obtained for all the points of the map (except at the border of the images, depending on the method used). They were analyzed using the following procedure: for each horizontal line ($x$=cst), i.e. for each spatial frequency, the best sine was extracted from experimental data points by using a minimization procedure. The relative difference between the extracted amplitude and the theoretical one is used as an error indicator, denoted *ErrA*. It can be presented as a function of the signal wavelength.

Two typical behaviors are shown in Fig. 3.

1. For the lowest signal wavelengths, the error is 100%. It decreases monotonically down to zero. In this case, the extraction algorithm always underestimates the amplitude. Because high frequencies are filtered out and lower frequencies are kept, the method behaves as a low-pass filter. Then, the slope of the second part of the curve – taken in a log-log representation – gives the order of the equivalent filter, i.e. its ability to reconstruct signals having a period higher that the cutting wavelength.

2. The error is 100% for the lowest signal wavelengths. *ErrA* decreases strongly, overestimates the amplitude and finally ends tending toward the theoretical amplitude. This behavior is well represented by a second order filter. It corresponds to SA, which has the particularity to extract directly the phase in the Fourier domain, with no explicit low-pass filter. SFT is working as well in the Fourier domain, but in a global manner, contrarily to SA, which is working with a small window, therefore close to the Nyquist frequency. Then, this behavior can be related to aliasing effects. Anyway, because high frequencies are poorly reconstructed, the method can still be characterized by a high cutting frequency.

{figure 3 might be around here}

Fig. 4 shows the systematic processing applied to *ErrA* vs spatial wavelength curves. In both cases, for the higher spatial frequencies, a first zone is described by a horizontal line characterized by *ErrA*=1. This indicates that such spatial wavelengths are strictly unreachable with the given image processing. The higher wavelength corresponding to this zone defines the ultimate spatial resolution (USR) of the given method: no localized information with a spatial wavelength lower than this USR can be detected. This gives a *qualitative* cutting frequency. In order to estimate a *quantitative* cutting frequency, it is possible to define another practical quantity corresponding to the maximal allowed error, for example 10%, denoted RS10%.

The last interesting quantity is the root mean square ($\sigma_U$) between the measured signal and the interpolated sine curve. It provides an indication on the random error that corrupts the signal. It can be plotted as a function of the spatial period.

{Figure 4 might be around here}

## 5 Results and discussion

*Pure sine wave*

First, the results are shown in Fig. 5 to 8. The optical model is a pure sine wave, encoded over 8 bits, without any noise and with a constant illumination (i.e. constant intensity offset and constant contrast). The grid step varies from 5 to 8 pixels per fringe. The results presented here are obtained according to the common practice of each laboratory.

Fig. 5 shows that spatial resolution at 10% error (SR10%) varies linearly with the ultimate spatial resolution (USR). Fig. 5 exhibits two groups of algorithm behaviors, indicated by slopes 1 and 2. This means that the low-pass filtering effects are not the same for all the demodulation techniques. The slope of SFT and SA algorithms is lower than the other ones: a local event would be rendered with a higher optical resolution with these methods compared to the other ones.

{Figure 5 might be around here}

In these two latter techniques, a rectangular window is directly applied in the Fourier domain, whereas, in the former ones, a rectangular (DIC) or triangular (SPS) window is applied in the real domain. The frequency filtering is different in these cases, as illustrated in Fig. 6: used in the Fourier domain, the rectangular window sharply rejects high frequencies; the two other windows present a continuous variation, indicating that the frequencies are biased by the filtering windows. In addition, the triangular window leads to a better high frequency rejection. Finally, image correlation techniques (using a rectangular window) behave roughly like spatial phase shifting techniques (using a triangular window.

{Figure 6 might be around here}

A better understanding on the USR can be found when it is plotted according to the characteristic typical length of the method, denoted *DimY* (Fig. 7), and which is defined by the dimension corresponding to the spectral bandwidth. It can be estimated using the following formula:

$$DimY = \frac{1}{f_{max} - f_{min}} \quad (29)$$

where $f_{max}$ and $f_{min}$ are the cut-off frequencies as defined in section 3.

In the cases of DIC, MPC and SPS, the cut-off frequencies are set to the first roots of the frequency spectrum (see Fig. 6). Note that the spectral length directly corresponds to the dimension of the subset in the case of a rectangular window (DIC, MPC), but is half the dimension subset in the case of the triangular window (SPS).

The USR varies roughly linearly with the subset dimension, with some visible scatter. The practice among the laboratories is very different: in some cases, the subset dimension is calculated from the grid step; in others, the subset dimension is kept fixed. As a matter of fact, the spatial resolution may vary without any change in the subset dimension (see DIC, MPC 8×2 and 24×1 or SA results), but only because of the effect of the encoding grid step. For the SFT technique, the vertical dimension is adapted according to the estimation of the frequency peak in the Fourier spectrum. In this case, the lower the central frequency, the smaller the filtering window and hence the larger $DimY$.

{Figure 7 might be around here}

The random errors induced by all demodulation techniques vary approximately from $10^{-2}$ to $10^{-3}$ pixel. There is no clear correlation between the random error and the USR, but with the subset area. Because, most of the methods are based on phase extraction, it seems reasonable to present the random error on the phase, $\sigma/P_{car}$, i.e. the phase in fraction of $2\pi$ (Fig. 8). Of course, this representation does not fit very well with the DIC procedure. Even if this representation is not usual in the analysis of DIC errors, it does make sense as the accuracy of DIC is dependent on the intensity of gray level gradients in the correlation window, which is directly related in the current case to the pattern size [5].

The global trend is a decrease of $\sigma/P_{car}$ with the subset area, showing a classic averaging effect; the mean slope is very close to the expected value of -0.5. SA and SFT results will be commented further. Both are spectral methods and both present higher noise level. In the first case, the noise on SA 16 can be explained by the small size of the window compared with the encoding grid step: the frequency detection is at the limits given by the Nyquist–Shannon sampling theorem, stating that the reconstructed signal from sampled data cannot have any frequencies higher than one-half of the sampling frequency. In order to have a larger window in the vertical direction, a size of 32x8 pixels gives the same subset area but allows us to obtain better results. Another error source is due to the interpolation procedure given in Eq. 20 induces some problems, and the best results are obtained in the case of $P_{car}=8$, because the interpolation has no effect ($N/P_{car}$ is close to an integer value) In the case of SFT, the larger the grid pitch, the smaller the frequency window in the Fourier domain. Consequently, some information contained in the rejected frequencies should have induced harmonics in the signal. Considering the estimation method used for the random noise, these harmonics should have increased its value.

Last, for DIC, the error increases strongly if the correlation window size is less than half the pitch of the grid. This is essentially due to the correlation criterion used in these simulations, which is the Zero Centered Normalized Cross Correlation. Indeed, this criterion does not take into account either the average grey level in the window or its standard deviation. So, a too small window is unable to provide reliable information for a one-to-one spatial positioning. A similar trend was already observed for random patterns [5].

{Figure 8 might be around here}

*Parametric study*

The effects of the random noise and the intensity shape on the resolution and the spatial resolution are examined here. No typical trend on the spatial resolution is detectable, but it is very different on the resolution.

In the case of a higher random noise in the intensity map, Fig. 9 shows that the random error is shifted over 1 decade compared to noiseless situation represented on Fig. 8. Indead the averaging effect is still the governing phenomenon. The noise level corresponds to the reasonable experimental values. Here, methods behave in the same way, with no significant changes.

The square grids give the worst results. In Fig. 10, the typical noise level is close to $1\times10^{-2}$, even if the signal-to-noise ratio SNR is 100. It is worth noting that using large subset areas is useless. This situation is particular: in practice, when using square grids, the image formation procedure tends to smooth out the grid boundaries. This case has to be considered as an asymptotic situation.

Methods behave in a different way: spectral analysis is more sensitive than phase-shifting methods. Among the phase shifting methods, even if the theoretical basis is the same, the LMPF implementation – which was the best for the previous situations, see Fig. 8 – is in this one the worst, showing a higher dependency on the signal quality; nevertheless, this trend remains small compared to the global variations of Fig. 10. Last, because it does not include any assumption on the intensity shape, DIC seems more efficient than the other analyses, which instrinsically assume a sinusoidal signal intensity shape.

{Figure 9 might be around here}

{Figure 10 might be around here}

# 6 Conclusions

In this paper, a new procedure has been proposed to assess the metrological performance of various image analysis techniques to extract displacement fields from deformed fringe patterns with a special emphasis both on the measurand resolution and the spatial resolution. The studied algorithms using a single pair of images are used in various set-ups in experimental mechanics, in particular grid techniques, but also specific high speed implementations of interferometry, holography, moiré, fringe projection…

The procedure proposed in this paper is based on the generation of a pair of synthetic images with a particular multi-frequency field. This field provides, within a single analysis, all the experimental errors on the spatial frequency of the measured signal. The result is given as a Bode diagram. This approach enables us to rigorously define the concept of the spatial resolution of a given image analysis technique. More precisely, two quantifications of the spatial resolutions have been proposed. First, the ultimate spatial resolution defines a detection level. Second, the obtained dependence of the measurement bias as a function of the frequency provides the spatial resolution associated with an error level which can be defined *a priori* (for instance 10%). In addition, as the procedure is rather fast, it is easy to investigate the influence of various image characteristics (pitch of the grid, image noise, shape of the fringe) and of the parameters of the image analysis techniques (window size, spectral bandwidth) on the quantities of interest.

This methodology has been applied to several image analysis techniques based either on fringe demodulation algorithms, on Fourier transform or on image correlation. These techniques are widely used in the laboratories of the co-authors. This paper shows that these techniques share some common features but have also some specificities:

- the bias varies as a power of the spatial wavelength of the measured signal;

- the exponent of the power law depends on the algorithm. Correlation and phase shifting algorithms exhibit a similar response whereas Fourier-based algorithms have a stronger low pass filtering effect;

- the spatial resolution mainly depends on the characteristic size of the windows (or the bandwidth used by the image processing). Both quantities are not exactly equal. Their ratio depends on the algorithm which is employed;

- in the case of a sinusoidal fringe pattern, the random error is strongly governed by the number of pixels involved in the local analysis. The global trend is a decrease of the error with the number of pixels. This corresponds to a classic averaging effect;

- this random error is very sensitive to the image noise level and to the profile of the simulated fringe. All algorithms behave similarly when sine-shaped fringes are considered, but significant differences are observed for rectangular profiles. This is due to the fact that some algorithms rely on an assumption on the fringe profile contrary to some others.

The current procedure shows that an optimization of the window size and the pitch can potentially be obtained for each method depending on the field under study, especially its displacement or strain gradients. This will be the aim of further studies on this approach.

The synthetic speckle images that were used to perform the simulations can be downloaded from the following website: http://gdr2519-images.mines-albi.fr/. The readers interested in assessing the metrological performance of their own software are invited to download these files, to use them as input data and to discuss matters related to their results with the authors.

# 7 Acknowledgements

The authors and all the participants of this benchmark are grateful to the CNRS for supporting this research.

# References

1. SPOTS Standardization project for optical techniques of strain measurements, EU contract G6RD-CT-2002-00856, see http://www.opticalstrain.org

2. Burguete R, Hack E, Patterson E, Siebert T, Whelan M (2010) Guidelines for the calibration of optical systems for strain measurements, Part I: calibration, see http://www.opticalstrain.org

3. Burguete R, Hack E, Patterson E, Siebert T, Whelan M (2010) Guidelines for the calibration of optical systems for strain measurements, Part II: validation, see http://www.opticalstrain.org

4. http://www.gdr2519.cnrs.fr/

5. Bornert M, Brémand F, Doumalin P, Dupré J-C, Fazzini M, Grédiac M, Hild F, Mistou S, Molimard J, Orteu J-J, Robert L, Surrel Y, Vacher P, Wattrisse B (2009) Assessment of Digital Image Correlation Measurement Errors: Methodology and Results. Experimental Mechanics 49 (3): 353-370

6. Kobayashi A S (ed) (1993) Handbook on Experimental Mechanics, second revised edition, Society for Experimental Mechanics. SEM and VCH, Bethel

7. Cloud G (1998) Optical Methods of Engineering Analysis. Cambridge University Press, Cambridge

8. Osten W (2000) Digital Processing and Evaluation of Fringe Patterns in Optical Metrology and Non-Destructive Testing. In: Laermann K-H (ed) Optical Methods in Experimental Solid Mechanics. Springer, CISM 403. Springer, pp 289-422

9. Surrel Y (2000) Fringe Analysis. In: Rastogi P K (ed) Photomechanics, Topics Appl. Phys. 77. Springer, pp 55-102

10. Surrel Y (1997) Additive noise effect in digital phase detection. Applied Optics 36(1): 271-276

11. Brémand F (1994) A phase unwrapping technique for object relief determination. Optics and Lasers in Engineering 21(1-2): 49-60

12. Sciammarella CA, Lamberti L, Sciammarella FM (2005) High-accuracy contouring using projection moiré. Optical Engineering 44(9): art. no. 093605

13. Y. surrel, N. Fournier, M. Grédiac, P.-A. Paris (1999) phase stepped deflectometry applied to shape measurement of bent plates, Experimental mechanics, 39(1): 66-70

14. Surrel Y (1996) Design of algorithms for phase measurements by the use of phase-stepping. Journal of Applied Optics 35(1): 51-60

15. Meinlschmidt P, Hinsch K, Sirohi R (Eds.) (1996). Selected Papers on Electronic Speckle Pattern Interferometry: Principle and Practice. Vol. MS 132. SPIE Optical Engineering Press

16. Rastogi PK (Eds.) (2000) Photomechanics (TAP/77). Springer, Berlin

17. Jacquot P, Fournier J-M (Eds.) (2000) Interferometry in Speckle Light. Springer, Berlin

18. Post D, Han B, Ifju P (1994) High sensitivity moire : experimental analysis for mechanics & materials (mechanical engineering series). Springer, Berlin

19. Morimoto Y, Nomura T, Fujigaki M, Yoneyama S, Takahashi I (2005) Deformation measurement by phase-shifting digital holography. Experimental Mechanics 45(1): 65-70

20. Picart P, Diouf B, Lolive E, Berthelot J-M (2004) Investigation of fracture mechanisms in resin concrete using spatially multiplexed digital Fresnel holograms. Optical Engineering 43(5): 1169-1176

21. Lee J-R, Molimard J, Vautrin A, Surrel Y (2004) Digital phase-shifting grating shearography for experimental analysis of fabric composites under tension. Composites Part A: Applied Science and Manufacturing 35 (7-8): 849-859

22. Dorrío BV, Fernández JL (1999) Phase-evaluation methods in whole-field optical measurement techniques. Measurement Science and Technology 10(3): R33-R55

23. Cordero, R., Molimard J., Martinez A. and Labbé F (2007) Uncertainty Analysis of Temporal Phase-Stepping Algorithms for Interferometry Optics Communications, 275: 144-155

24. Takeda M, Ina H, Kobayashi S (1982) Fourier-transform method of fringe-pattern analysis for computer-based topography and interferometry. J. Opt. Soc. Am. 72 (1): 156-160

25. Desse J-M, Picart P, Tankam P (2008) Digital three-color holographic interferometry for flow analysis. Optics Express 16(8): 5471-5480

26. Xianyu Su, Wenjing Chen (2001), Fourier transform profilometry:: a review, Optics and Lasers in Engineering, 35(5), 263–284

27. Rajaona R D, Sulmont P (1985) A method of spectral analysis applied to periodic and pseudoperiodic signals. Journal of computational Physics 61(1): 186-193

28. Dupré J-C, Brémand F, Lagarde A (1993) Numerical spectral analysis of a grid: Application to strain measurements. Optics and Lasers in Engineering 18(3): 159-172

29. Doumalin P, Bornert M (2000) Micromechanical Applications of Digital Image Correlation Techniques. In: Jacquot P and Fournier J-M (eds) Interferometry in Speckle Light. Springer, Lausanne, pp 67-74

10. Cottron M, Brémand F, Lagarde A (1992) Non-contact and non-disturbing local strain measurement methods Part II: Application. European Journal of Mechanics 11(3): 367-379

31. Pannier Y, Avril S, Rotinat R, Pierron F (2006) Identification of Elasto-Plastic Constitutive Parameters from Statically Undetermined Tests Using the Virtual Fields Method. Experimental Mechanics 46 (6) : 735-755

32. Cordero R R, Molimard J, Labbé F, Martínez A (2008) Strain maps obtained by phase-shifting interferometry: An uncertainty analysis. Optics Communications 281(8): 2195-2206

33. Badulescu C, Grédiac M, Mathias J-D, Roux D (2009) A Procedure for Accurate One-Dimensional Strain Measurement Using the Grid Method. Experimental Mechanics 49(6): 841-854

34. Avril S., Vautrin A., Surrel Y. (2004) Grid Method: Application to the Characterization of Cracks, Experimental Mechanics 44 (1), 37-43

35. Moulart R, Rotinat R, Pierron F (2009) Full-field evaluation of the onset of microplasticity in a steel specimen. Mechanics of Materials 41(11): 1207-1222

36. Pastor M.L., Balandraud X., Robert J.L., Grédiac M. (2009) Lifetime prediction of aluminium structures reinforced with composite patches, International Journal of Fatigue, 31 (5), 850–858

37. Robin E, Valle V (2004) Phase demodulation from a single fringe pattern based on a correlation technique. Journal of Applied Optics 43(22): 4355-4361

38. Robin E, Valle V, Brémand F (2005) Phase demodulation method from a single fringe pattern based on correlation with a polynomial form. Journal of Applied Optics 44(34): 7261-7269

37. Héripré E, Dexet M, Crépin J, Gélébart L, Roos A, Bornert M, Caldemaison D (2207) Coupling between experimental measurements and polycrystal finite element calculations for micromechanical study of metallic materials. International Journal of Plasticity 23(9): 1512-1539

# Tables

|  | Case 1 | Case 2 | Case 3 | Case 4 |
|---|---|---|---|---|
| *Variable data* | | | | |
| $P_{car}$ (pixel) | 5, 6, 7, 8 | 5, 6, 7, 8 | 5, 6, 7, 8 | 5, 6, 7, 8 |
| *frng* function | sine | sine | rectangular | Rectangular |
| SNR | 100 | 30 | 100 | 30 |
| *Fixed data* | | | | |
| $x_{max}$ (pixel) | | 512 | | |
| $y_{max}$ (pixel) | | 1024 | | |
| $P$ (pixel) | | 5 | | |
| $A_U$ (pixel) | | 0.3 | | |
| Over-sampling | | 4×4 | | |
| $I_0$ (GL) | | 128 | | |
| $\gamma$ (GL) | | 0.59 | | |
| Encoding depth (bits) | | 8 | | |

Table 1. Parameters used for the round robin test

|  | Case 5 | Case 6 | Case 7 |
|---|---|---|---|
| $P_{car}$ (pixel) | 5, 6, 7, 8 | 5, 6, 7, 8 | 5 |
| *frng* function | sine | sine | sine |
| SNR | 100 | 30 | 100 |
| $x_{max}$ (pixel) | 512 | 512 | 512 |
| $y_{max}$ (pixel) | 1024 | 1024 | 1024 |
| $P$ (pixel) | 5 | 5 | 5 |
| $A_U$ (pixel) | 0.15 | 0.15 | 0.3 |
| Over-sampling | 4×4 | 4×4 | 1×1 |
| $I_0$ (GL) | 128 | 128 | 128 |
| $\gamma$ (GL) | 0.59 | 0.59 | 0.59 |
| Encoding depth (bits) | 8 | 8 | 8 |

Table 2. Parameters used for the check tests

# Figures

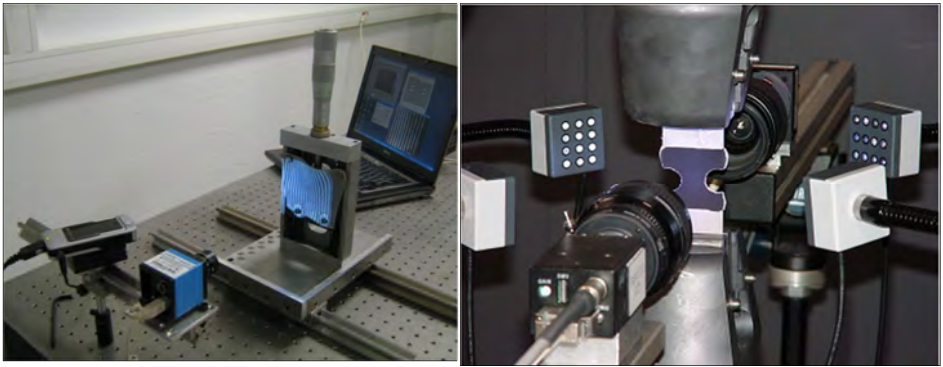

(a)                                                  (b)

**Fig.1.** Fringe generation in the case (a) of projected fringes (courtesy of EMSE) (b) of the grid method (courtesy of Arts et Métiers ParisTech)

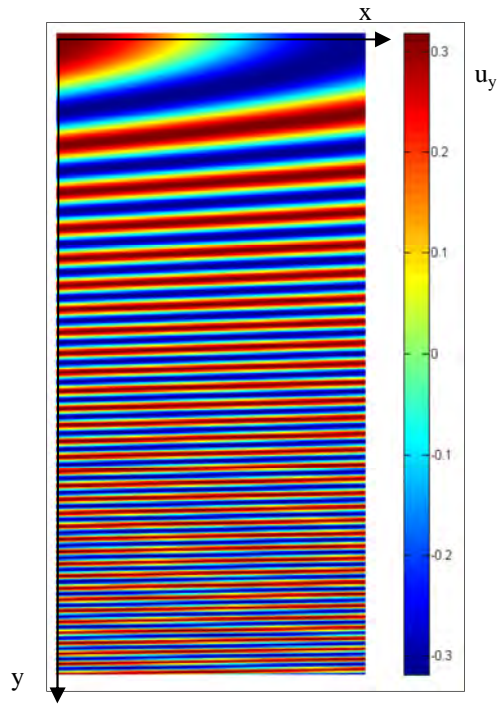

**Fig.2.** Example of displacement field ($A_U$=0.3, $P$=5, $y_{max}$=1024 and $x_{max}$=512 in pixels)

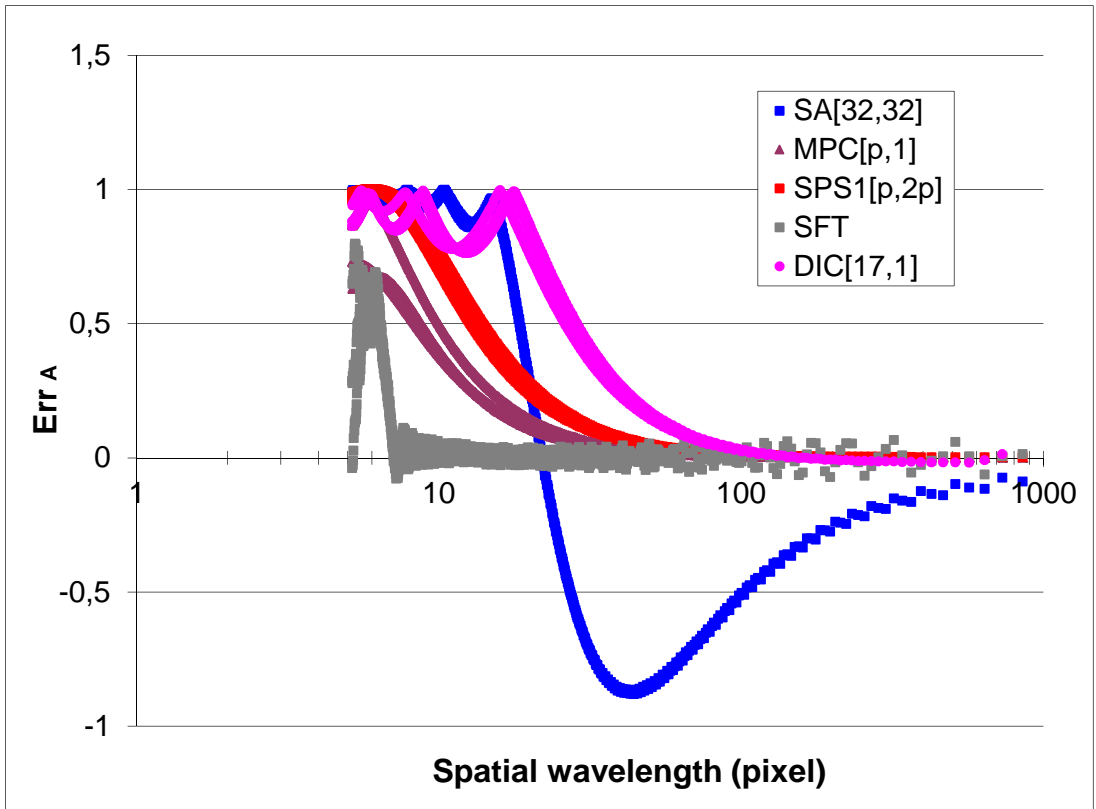

**Fig.3.** Example of typical curves *ErrA* (in pixels) vs spatial wavelength.(p=6 pixels)

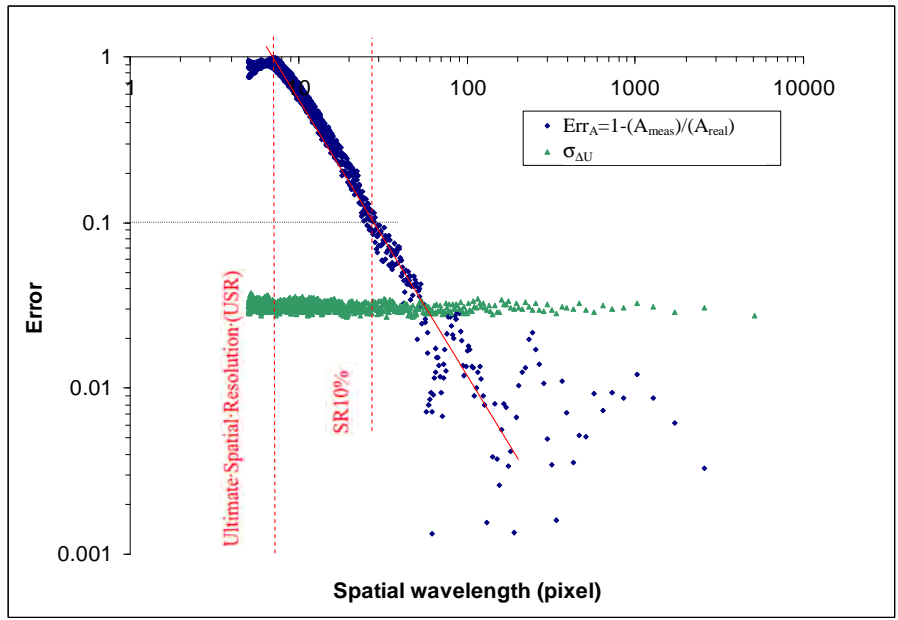

**Fig.4.** Analysis of curves *ErrA* or $\sigma_U$ vs spatial wavelength.

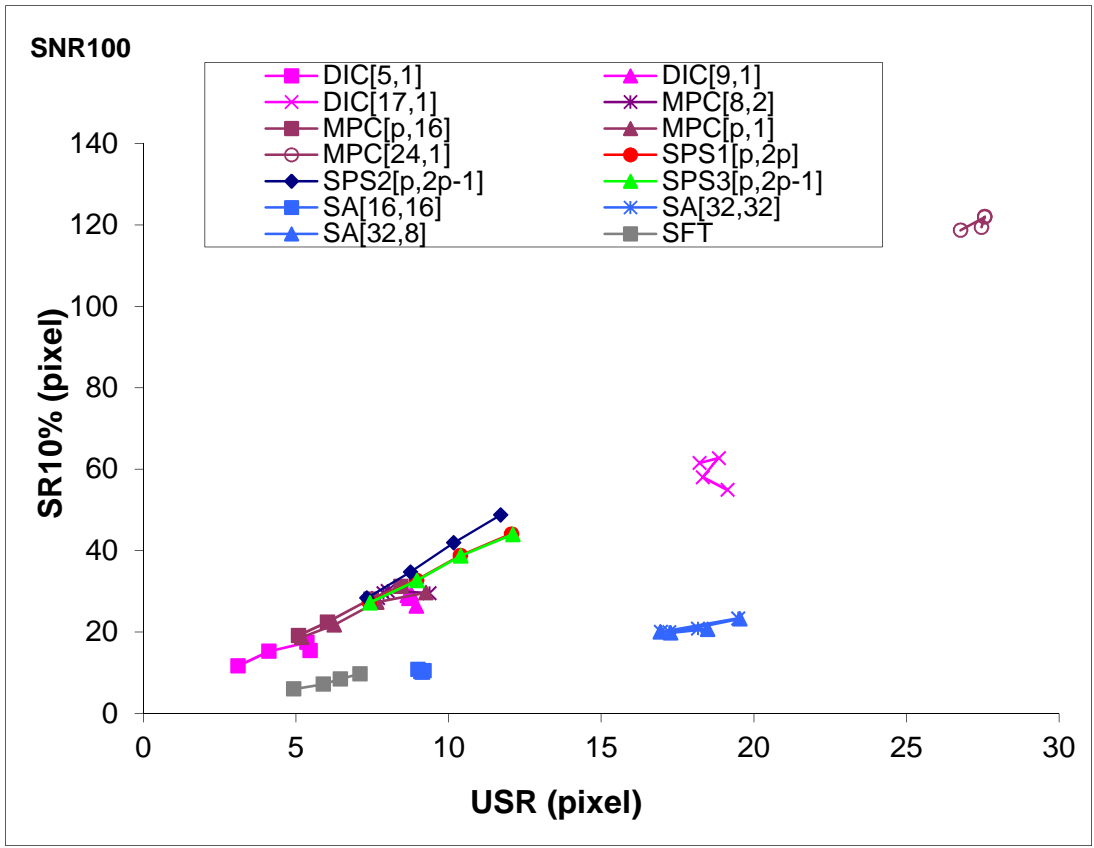

**Fig. 5.** 10% Spatial Resolution vs ultimate Spatial Resolution

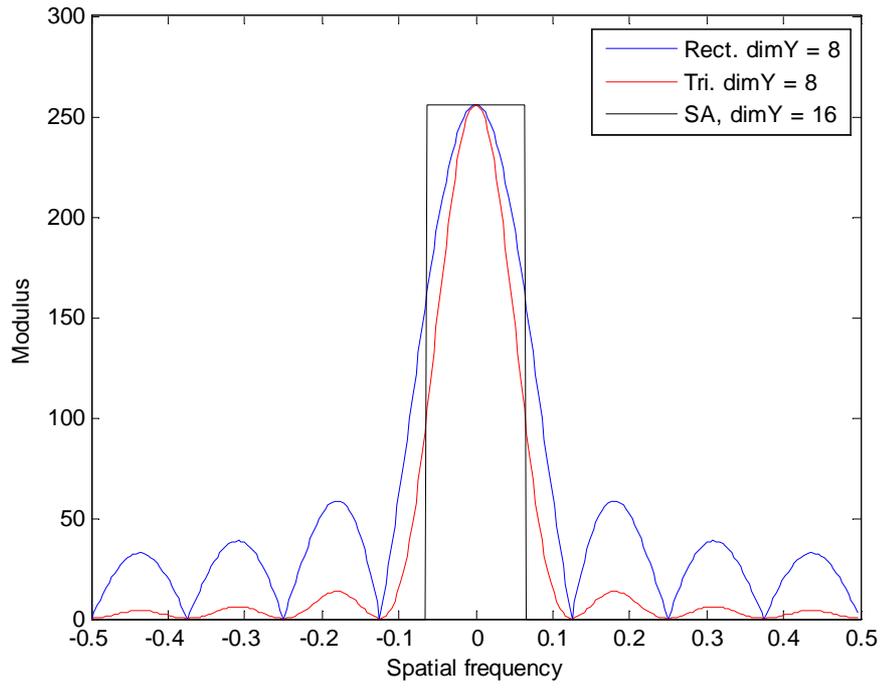

**Fig. 6.** Fourier Transform of the different windows, $P_{car}$=8, vertical subset dimension 8 (rectangular), 15 (triangular), 32 (spectral analysis).

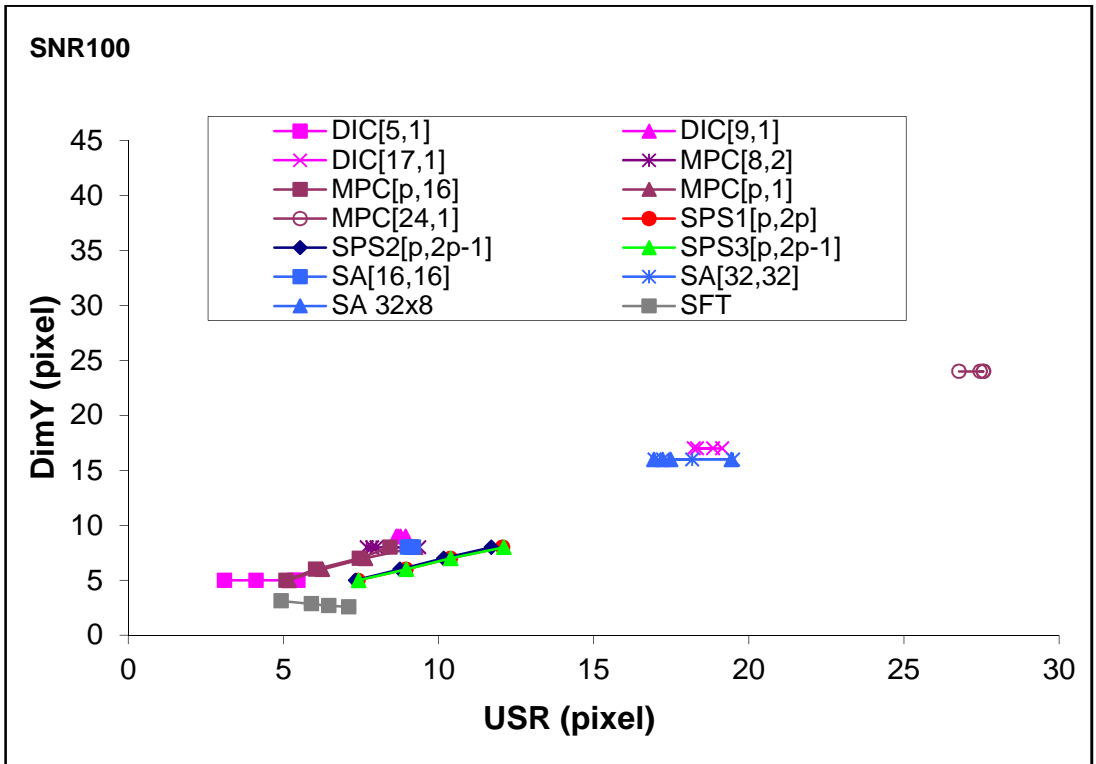

**Fig. 7.** Ultimate spatial resolution vs vertical subset dimension

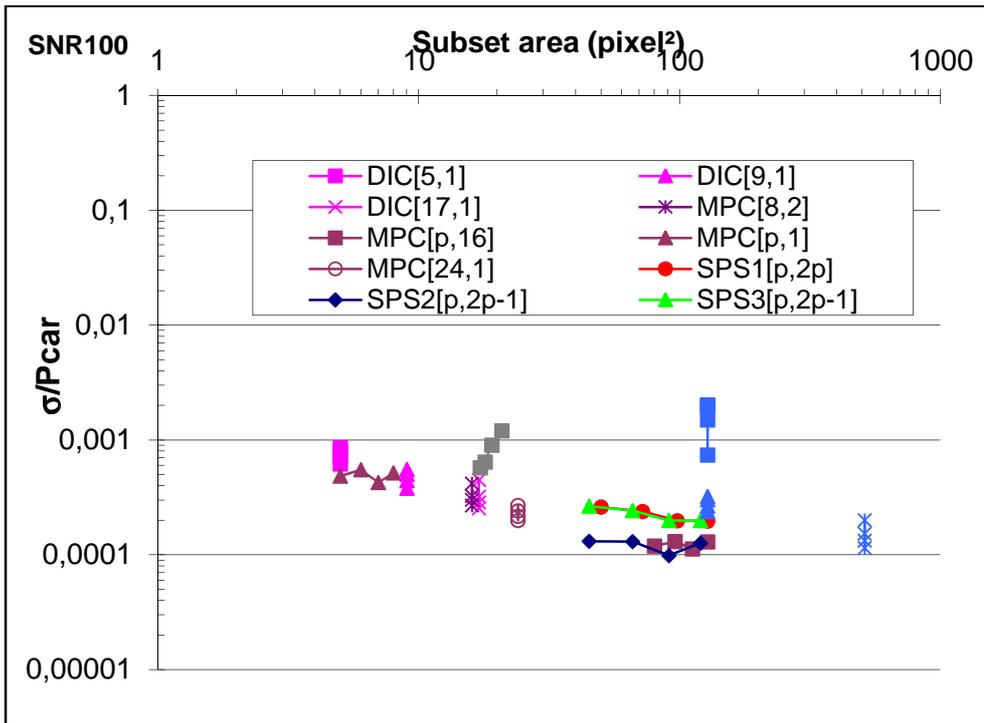

Fig. 8. Random error vs subset dimension, SNR=100

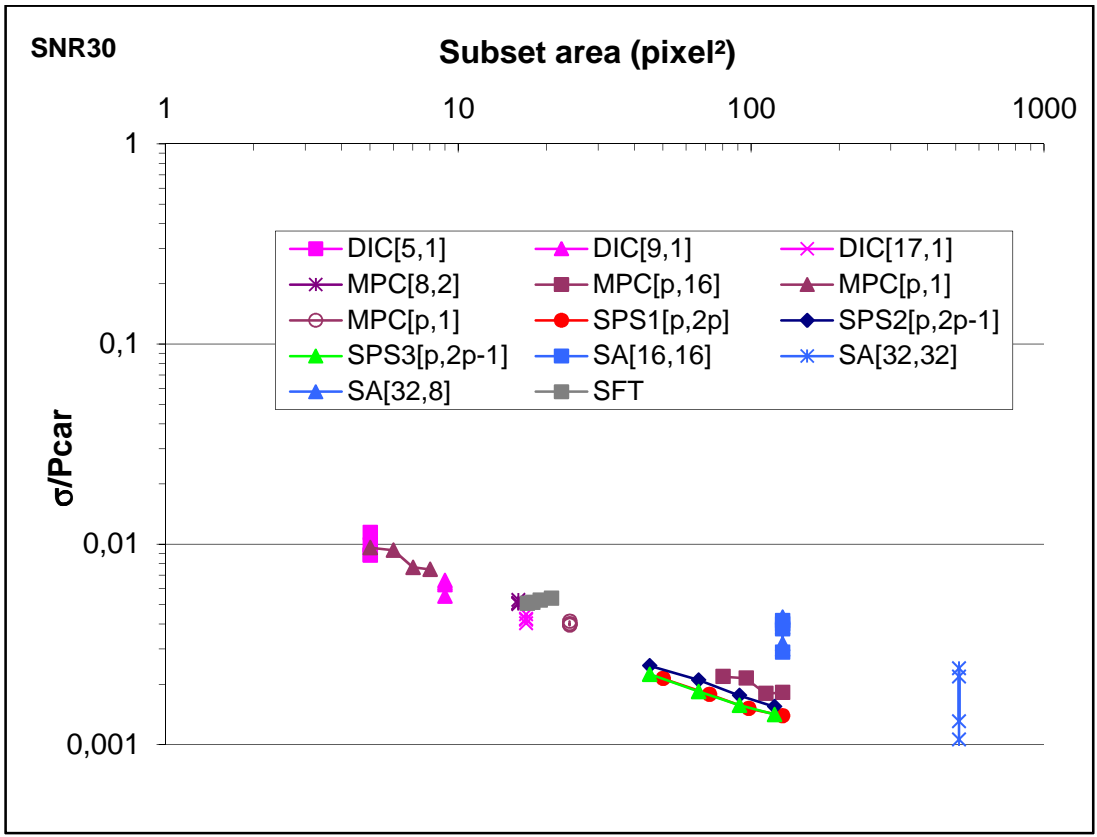

**Fig. 9.** Random error vs subset dimension, SNR=30

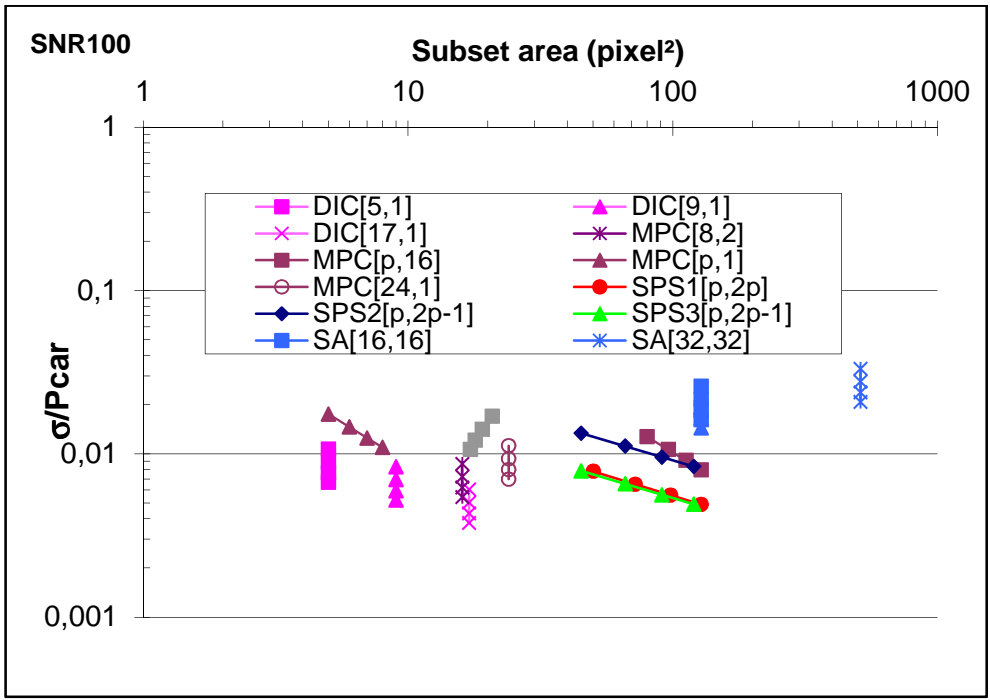

**Fig. 10.** Random error vs subset dimension, square grids